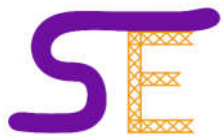

# ComContA: Content-related Analysis of Communication in Groupchats of Software Development Teams

*Work Package WP5*
**Synthesis of Insights and Implementation**

**SEnti-Analyzer: Joint Sentiment Analysis For Text-Based and Verbal Communication in Software Projects**
– Technical Report –

authored by
**Marc Herrmann, Martin Obaidi, and Jil Klünder**

*Funding Instrument:*   Leibniz University Hannover
*Funding:*   Leibniz Young Investigator Grant
*Project lead:*   Dr. Jil Klünder
*Project Duration:*   2020 – 2022

# Deliverable Information

| Document Administrative Information | |
|---|---|
| Project Acronym: | ComContA |
| Project Number: | 85430128 |
| Deliverable Number: | D5.1 |
| Deliverable Full Title: | SEnti-Analyzer: Joint Sentiment Analysis For Text-Based and Verbal Communication in Software Projects |
| Deliverable Short Title: | SEnti-Analyzer |
| Authored by: | Marc Herrmann, Martin Obaidi, and Jil Klünder |
| Report Version: | v1.0 |
| Report Submission Date: | June 22, 2022 |
| Dissemination Level: | PU |
| Nature: | Technical Report |

# Change Log

| Date | Version | Author/Editor | Summary of Changes Made |
|---|---|---|---|
| June 22, 2022 | v1.0 | Jil Klünder | Initial version |

| Dissemination Level | | |
|---|---|---|
| PU | Public | ✓ |
| PP | Restricted to other programme participants (including the Commission Services) | |
| RE | Restricted to a group specified by the Consortium (including the Commission Services) | |
| CO | Confidential, only for members of the Consortium (including the Commission Services) | |

# Summary

**Context:** Social aspects in software development teams are of particular importance for a successful project closure. To analyze sentiments in software projects, there are several tools and approaches available. These tools analyze text-based communication based on the used words to predict whether they appear to be positive, negative, or neutral for the receiver of the message.

**Research Project ComContA:** In the research project ComContA, we investigate so-called sentiment analysis in software engineering striving to analyze the content of text-based communication in development teams with regard to the statement's polarity. That is, we tend to analyze whether the communication appears to be adequate (i.e., positive or neutral) or negative and requires improvement.

**The SEnti-Analyzer:** In a workshop paper, we presented a tool called *SEnti-Analyzer* that allows to apply sentiment analysis to verbal communication in meetings of software projects. We evaluated the tool in a student software project for its applicability, but not for its accuracy. In this technical report, we present the extended functionalities of the *SEnti-Analyzer* by also allowing the analysis of text-based communication (which is so far the main use case of sentiment analysis), we improve the prediction of the tool by including established sentiment analysis tools, and we evaluate the tool with respect to its accuracy.

**Evaluation:** We evaluate the tool by comparing (1) the prediction of the *SEnti-Analyzer* to (2) pre-labeled established data sets used for sentiment analysis in software engineering and to (3) to perceptions of computer scientists. Our results indicate that in almost all cases (91 of 96) at least two of the three votes coincide, but in only about half of the cases all three votes coincide.

**Next Steps and Future Research Directions:** Our results raise the question of the "ultimate truth" of sentiment analysis outcomes: What do we want to predict with sentiment analysis tools? The pre-defined labels of established data sets? The perception of computer scientists? Or the perception of single computer scientists which appears to be the most meaningful objective? But how can this be achieved?

# Structure of the Technical Report

This document presents information on the SEnti-Analyzer, its applicability, and its functionality. Section 1 presents the motivation and background of this technical report. In Section 2, we locate this research in contrast to existing literature and provide background information. Section 3 summarizes information on the software in its current form, as well as the techniques implemented to pre-process and classify the input data. We describe how we evaluated the software in Section 4 and present the results in Section 5. Finally, we discuss our results in Section 6 and conclude our work in Section 7.

# Table of Contents

# 1   Introduction

As modern software projects often require team work, social aspects of the development team are of particular importance [31]. Team-internal conflicts, frustrated team members, or other difficulties can endanger project success [15, 10]. However, observing aspects like communication behavior, the workload, or conflicts can help project managers to lead the project to success [25, 29]. Emotions or sentiments are one of the important social aspects [30]. For example, happiness has been shown to positively influence productivity and the ability to solve problems, which are both relevant aspects for a successful project closure [14, 37]. Consequently, project leaders and managers are interested in being aware of the temporary emotional shade in the team, which we refer to as mood. In the last years, so-called sentiment analysis has gained increasing relevance in research and practice [43]. The sentiment analysis tools analyze text-based communication with respect to the transported polarity or other sentiments [4]. So far, sentiment analysis has been frequently applied in software engineering to text-based communication in, e.g., group chats, JIRA, or other platforms [43].

In a workshop paper [18], we presented a concept to apply sentiment analysis to verbal communication in meetings of software projects. This application to meetings was motivated by the huge amount of communication in software project meetings, despite the frequent use of text-based communication channels. Although research has found meeting analysis to be of particular importance [48, 28], there was, to the best of our knowledge, so far no attempt to apply sentiment analysis in meetings. In a first attempt to close this gap, we developed a tool, the so-called *SEnti-Analyzer*, which transcribes the statements made in a meeting in real-time and applies existing sentiment analysis tools to the statements. Afterwards, the team receives feedback about the number of positive, negative, or neutral statements in the meeting. In the workshop paper [18], we applied the *SEnti-Analyzer* to a student software project and showed that

(1) Sentiment analysis can be applied to meetings, and

(2) The application of sentiment analysis to verbal communication is as meaningful as the application to textual communication.

In the paper [18], we showed the applicability of the tool, and, hence, that the workflow of our tool leads to results. However, we did not analyze the results for correctness. In this technical report, we strive to present the current state of the software as well as its applicability. We describe how we extended the functionality of the tool to be also applicable to text-based communication. This extension allowed us to test the predictions of the tool against existing established data sets [41, 34], that are pre-labeled and often used to train sentiment analysis tools. This functional extension brings us closer to the goal of developing a universal tool that can be used for multiple application scenarios (live audio recording, audio recording, text-based communication data) and is thus more practical. In addition, we use multiple lexicon-based sentiment analysis tools in an ensemble to ensure robust sentiment analysis in meetings. This is because it has been observed that lexicon-based tools such as SentiStrength or SentiStrength-SE, which do not require pre-training, perform well across multiple domains (such as GitHub or Stack Overflow data sets), while pre-trained machine learning tools perform worse on other, unknown data sets (e.g. [7, 52, 23, 41]). Moreover, due to their functionality, these lexicon-based tools can perform a fast sentiment analysis and are therefore well suited for a real-time analysis as proposed by [49]. Since, to the best of our knowledge, there is no data set consisting of communication data in meetings in the context of software engineering and thus pre-training for this domain is not possible, we hope to achieve a robust sentiment analysis with the choice of lexicon-based tools.

In addition to the functional extension, we compare the prediction of the tool with insights from a survey [19] in which computer scientists reported on their perception of statements (i.e., whether they appear to be positive, negative, or neutral). These two comparisons allow us to evaluate the accuracy of the *SEnti-Analyzer* and show that in almost all cases at least two of the three votes (*SEnti-Analyzer*, labels in the data set, and computer scientists) coincide, but in only 46 of the cases, all three votes coincide. Interestingly for 91 out of the 96 overall statements there are always at least two out of three who agree on the sentiment polarity label.

This insight points to the necessity to clarify the aspired outcome of sentiment analysis tools. Shall they predict the labels in a data set? The perception of a typical computer scientist? Or is there a need to adjust the tools to a specific team? The last option would lead to calibrated and, hence, more specific results that fit the team which uses the tool. Therefore, this option appears to be meaningful, but requires a lot of research to achieve this goal.

*Context.* The initial version of the SEnti-Analyzer was presented in the workhop paper "From Textual to Verbal Communication: Towards Applying Sentiment Analysis to a Software Project Meeting" [18] at the Fourth International Workshop on Affective Computing in Requirements Engineering in 2021. This technical report is based on this initial paper, but extends it as follows (according to the appearance in the paper):

- We extended the related work (Section 2) to fit the (new) scope of this paper.

- We added Section 3 describing the *SEnti-Analyzer* with the initial and the new functionalities (Section 3.1).

- In the study (Section 4), we added two new research questions to assess the accuracy of our approach.

- We compare the outcome of the tool to pre-labeled data sets (Section 5.2) and to perceptions of computer scientists (Section 5.3).

## 2 Background and Related Work

Meeting analysis and sentiment analysis have both been frequently applied to software projects. In the following, we present research that is related to the work presented in this paper.

### 2.1 Meeting Analysis in Software Engineering

Klünder et al. [28] elaborate the coding scheme *act4teams-SHORT*, which they derived from an established interaction analysis scheme in psychology. Using *act4teams-SHORT*, statements in a meeting can be categorized in one of eleven different categories such as "naming problems" or "giving information". According to the results of Klünder et al. [28], using this coding scheme and analyzing the resulting interactions in each category help identifying possible problematic behavior. Resolving this kind of behavior at early stages of a software project can lead to better overall team performance and project success [28]. This categorization of high level interaction analysis can be traced back to basic low level sentiment analysis which finds more and more applications recently.

### 2.2 Sentiment Analysis

There are numerous general sentiment analysis tools available some of which are related to the software engineering domain [6, 3, 21], while others are not. Thelwall et al. [53] developed *SentiStrength*, a sentiment analysis tool that work on sentiment polarity lexica interchangeable for different languages or purposes. For the German languages available sentiment analysis tools include *BertDE* by Guhr et al. [16] and *GerVADER* by Tymann et al. [54].

Tools for languages differing from English are still rare. Related to software engineering, Klünder et al. [26] developed a classifier for German text messages from group chats of development teams that maps the input data to the polarity of the message, i.e., *positive*, *negative*, or *neutral*. This classifier is based on a trained classification model and defines a key part of the *SEnti-Analyzer* which we present in this paper. Calefato et al. [6] present their tool *Senti4SD* to provide a sentiment analysis tool trained on the software engineering domain (using data from *Stack Overflow*), which decreases the risk of misclassification due to the associated terminology. A similar approach is used by Islam et al. [22], using *JIRA* issue comments for training their tool named *SentiStrength-SE*, and Ahmed et al. [3], who use code review comments for training their tool *SentiCR*.

### 2.3 Data Sets for Sentiment Analysis

To verify the performance of sentiment analysis tools, pre-labeled data sets are commonly used. Lin et al. [34] provide a data set[1] from 1500 collected discussions on Stack Overflow tagged with the term "Java". Since no use of any guidelines is mentioned in the paper, we assume the authors labeled the statements ad hoc. Novielli et al. [41] present their gold standard data set[2] crawled from the collaborative development website GitHub, containing over 7000 manually annotated statements. They first assigned emotions to each sentence using an emotion framework by Shaver et al. [50] and later labeled polarities based on these emotions [41].

---

[1] The Stack Overflow data set from Lin et al. [34] is available on GitHub.
[2] The GitHub gold standard data set from Novielli et al. [41] is available on Figshare.

## 2.4 Majority Voting for Sentiment Analysis

The concept of majority voting by multiple classifiers has already been applied in the field of software engineering.

Moustafa et al. [38] developed a software bug prediction model using weighted majority voting techniques. They applied their model using different sets of software metrics for the classification. Afterwards, they tested their approach on data sets of different sizes. They found that ensembles tend to be more accurate than their base classifiers.

Onan et al. [44] wanted to enhance the predictive performance of sentiment classification. Therefore, they developed a paradigm of a weighted majority voting scheme to assign appropriate weight values to classifiers (e.g. logistic regression) and each output class based on the predictive performance of classification algorithms. Their experimental analysis included classification tasks like sentiment analysis, software defect prediction, and spam filtering. They concluded that their classification scheme can predict better than conventional ensemble learning methods.

Malhotra and Khanna [36] used different majority voting classifiers to predict change prone classes of a software. They used a set of Particle Swarm Optimization for classification and used their prediction in a voting method. The resulting model, which had included weighted fitness values, had an improved accuracy.

Liu et al. [35] used several classifiers like Baseline Classifier, Validation Classifier, and Validation-and-Voting Classifier, which uses a majority voting approach, to train a software quality model. They goal was to provide better generalization and more robust software quality models. Among these classifiers, the Validation-and-Voting Classifier performed best.

Liaqat et al. [33] has developed Majority Voting Goal Based (MVGB) technique for requirements prioritization. This is designed to allow the selection of requirements that satisfy the specified goals in situations of deadlock in requirements prioritization during the software development phase. The MVGB was compared with other techniques such as Hierarchy Cumulative Voting or Analytical Hierarchy Process and performed best.

## 2.5 Speech Recognition

Besides sentiment analysis, our approach is related to previous research in speech recognition. Agarwal and Zesch [2] present their approach in training a German-language model for the *Mozilla DeepSpeech* framework, which also constitutes the foundation of our speech recognition. The framework provides a transcript that can be used as input for existing sentiment analysis tools.

Speech recognition has also been applied to meetings in software engineering with another focus: Gall and Berenbach [13] present a framework recording requirements elicitation meetings on video, thereby collecting relevant information raised by stakeholders. Shakeri et al. [1] also strive to extract relevant information presented in elicitation meetings. Their tool *ELICA* collects knowledge and information related to requirements. This way, it helps analyzing the meeting outcome.

However, according to an systematic literature review conducted by Obaidi and Klünder [43], these sentiment analysis tools and other tools they found analyze only text-based communication and cannot analyze audio recordings.

In this paper, we combine the approaches of sentiment analysis tools with automatic speech recognition to analyze verbal and textual team communication, where direct feedback of the system of multiple sentiment analysis tools used is generated after the course of a meeting. For the fast sentiment analysis tools incorporated in the *SEnti-Analyzer* feedback can even be generated alongside the course of the meeting after each sentence spoken in real-time. This feedback can be used by project managers, software process engineers, and the like.

Although both sentiment analysis and meeting analysis have been proven to be beneficial for software projects, to the best of our knowledge sentiment analysis has not yet been used for meeting analysis.

The approach presented in this paper strives to combine the core topics of all of these papers. To the best of our knowledge, this is the first time of combining speech recognition for meeting analysis with sentiment analysis in software engineering. However, our work is based on existing research, most importantly, we use established sentiment analysis tools (both for software engineering and in general).

# 3 SEnti-Analyzer

The motivation for the development of the *SEnti-Analyzer* was to obtain proof of concept that a (real-time) end-to-end speech to sentiment polarity classification approach is feasible [18]. There are several existing sentiment analysis tools including ones specifically designed to be applied in software engineering domains such as *JIRA* [22], *GitHub* [3], and *Stack Overflow* [6]. However, all those tools require some technical expertise in data processing to obtain results for given data. Furthermore, we wanted to focus on analyzing the verbal communication that takes place within software projects such as meetings [18], while available tools are only able to analyze textual inputs [43].

In the following, we present details on how we obtain and (pre-)process the user data in the *SEnti-Analyzer*. In Section 3.1 we will highlight the improvements made to the *SEnti-Analyzer* since the initial presentation in the previous paper [18]. Figure 1 shows a simplified overview of the data processing steps in the version of the *SEnti-Analyzer*, which we will break down in the following subsections.

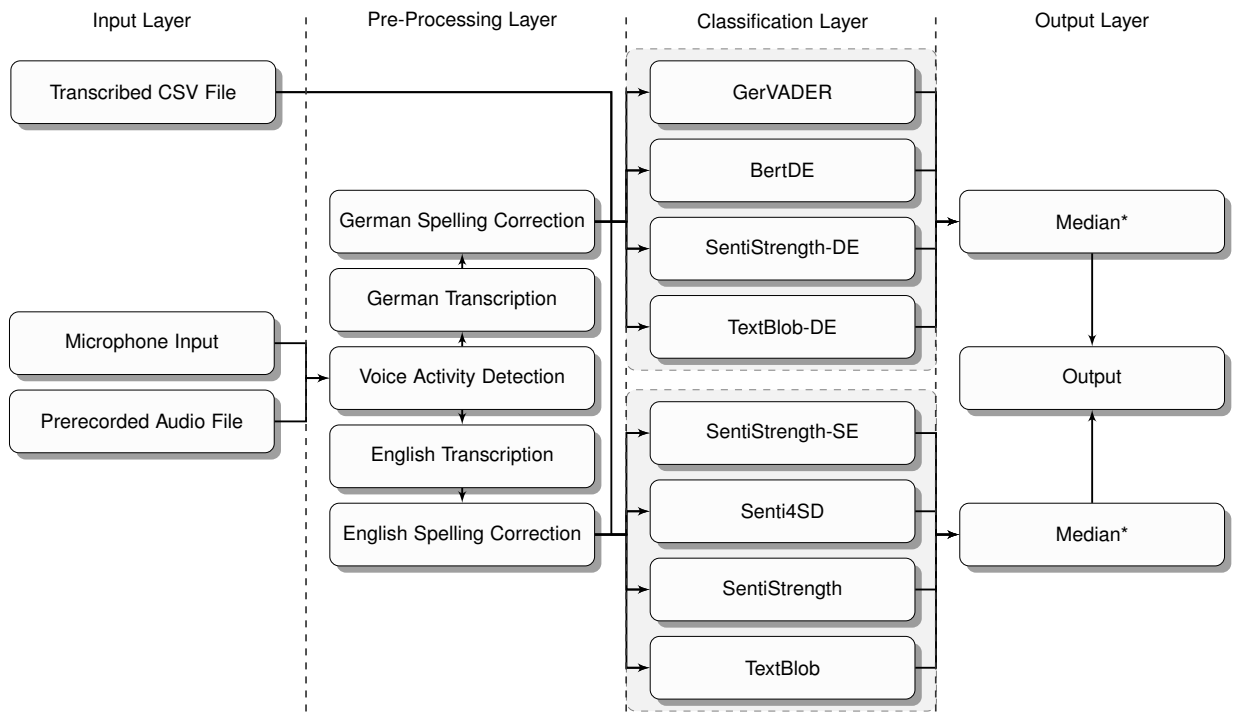

Figure 1: Processing pipeline of the SEnti-Analyzer for both German and English audio or CSV files

## 3.1 Improvements and changes made to the SEnti-Analyzer

The initial approach for the *SEnti-Analyzer* was to feed the users microphone input into our software, e.g., the microphone of a laptop placed in the middle of a conference table during the meeting [18]. We use voice activity detection and the *Mozilla Deepspeech* speech engine alongside with German language models [2] to transcribe utterances. After stopping the recording, the application of natural language processing to the transcript gets started to inherit a set of features and metrics required for the classification. Once completed, the collected statements and corresponding metrics are fed into the sentiment analysis tool provided by Klünder et al. [26] which interprets the results. Finally, an output of classified interactions is presented to the user, e.g., by calculating the total and relative proportions of each category (negative, neutral, and positive). Thus far, the following further developments (visualized in Figure 1) have been applied to the approach:

- The original sentiment analysis tool provided by Klünder et al. [26] was replaced with a set of four German sentiment analysis classification tools, for performance reasons.
- The option for English transcription and classification has been added alongside with English language models and four English sentiment analysis classification tools to be able to process English and German input data.
- The predictions of each of the four classification tools are combined to an overall outputted label following a given set of rules outlined in Section 3.5.
- After the transcription of statements, a spelling correction for English and German (capitalization in particular) respectively has been implemented before classification to improve the reliability of the results.
- The option to analyze a given set of pre-transcribed statements in the form of a CSV file has been added to allow the processing of both verbal and text-based communication.

The replacement of the initially used sentiment analysis tool provided by Klünder et al. [26] was mainly motivated by the size of the data set used to train this tool and the resulting small reliability [26].

## 3.2 Data Input

Three options are possible for inputting data into the *SEnti-Analyzer* as illustrated in the input layer section of Figure 1. The first and simplest one is to analyze a given set of statements which may have been pre-transcribed or come from a textual data sources. The set of statements must be obtained within a CSV File containing a "Text" named column containing one statement per row. Each statement then gets directly passed into the classification layer described in Section 3.4, as no pre-processing is needed for textual data. The second, and probably most interesting mode of operation is to directly input live audio into the *SEnti-Analyzer* via a connected or build-in microphone of the device (currently desktop devices are supported on all common operating systems). The raw audio data stream gets directly passed into the pre-processing layer described in subsection 3.3 to be prepared for the classification. Lastly, as an alternative for using the live microphone audio, prerecorded audio files can be inputted into the *SEnti-Analyzer*. This allows for asynchronous operation as well as preparation (i.e. clean up of noise) of the audio data. The only limitation is that audio files can be inputted in the Waveform format with a mono audio track and a sample rate of 16kHz, due to the text to speech engine we use. Otherwise, the prerecorded audio gets processed in the same way as the live microphone audio, by directly feeding the audio stream into the pre-processing layer.

## 3.3 Data Pre-Processing

The raw audio data stream received from the input layer either by the users' microphone input or a prerecorded audio file is prepared for the classification by different sentiment analysis tools in the pre-processing layer. The first problem we face is that we cannot transcribe our incoming audio stream directly because this would give us only a single statement without any kind of separation. The most common approach for splitting up verbal communication is known as voice activity detection (VAD) [46], which splits our input stream into utterances between every two moments of silence over the course of the whole audio stream. For this task we opted for the Google WebRTC [51] Voice Activity Detection as being performant and free to use. Each of the utterances then gets transcribed as soon as the speaker stops speaking and the transcribed statement is printed directly within the tool alongside the course of the conversation. For converting the speech to text not many of the readily available options were qualified, as we needed software that can perform speech to text transcription offline and for free, while most available services are paid APIs. We ultimately opted for the *Mozilla DeepSpeech* engine as it seemed to promise the best results and met our criteria. For the English transcription, we use the newest available language model available directly by *Mozilla*. For the German transcription, we use the language model published by Agarwal and Zesch [2]. Whether the English or German language model is used for transcription depends on the fact whether the user entered the option specifying that the data is German beforehand, otherwise the *SEnti-Analyzer* will by default use the English mode of operation. The same goes for the spelling correction which is applied to each of the statements obtained in the transcription step beforehand. While both of the language models for *DeepSpeech* already use n-gram models for the transcription, internally checking everything afterwards probabilistic NLP is beneficial. After the spelling correction the statements get passed into the classification layer described in the next subsection.

## 3.4 Classification

To classify our audio data, which has been pre-processed to textual data in the previous layer, we use a set of four sentiment analysis tools for each language. The statements are passed to the English sentiment analysis tools or to the German sentiment analysis tools depending on the pre-selected language of the data. The sentiment polarity predictions are then calculated for each of the four tools individually and stored in a data frame alongside with each statement[3]. The English sentiment analysis tools consist of:

- *Senti4SD* [6]
- *SentiStrength-SE* [21]
- *SentiStrength* [53]
- *TextBlob* ⇒ real time feedback

Two of the included English tools (Senti4SD and SentiStrength-SE) are specific to the software engineering domain. The fact that the other two tools are not specifically pre-trained to the software engineering domain is negligible for us due to the focus on meeting analysis which is mostly natural language. For the German language, we are not aware of even a single sentiment analysis tool specific to the software engineering domain available (despite the tool by Klünder et al. [26] which has a rather low reliability). Therefore, the German sentiment analysis tools used in the *SEnti-Analyzer* consist of:

- *BertDE* [16]
- *GerVADER* [54]
- *SentiStrength-DE*
- *TextBlob-DE* ⇒ real time feedback

---
[3]Note that this is bypassed for the fastest out of the 4 tools for each language respectively to provide the real-time feedback immediately alongside with the transcript itself).

The resulting sentiment polarity labels (negative, neutral, positive) or values in the interval [-1, 1] are added to a data frame alongside with the collected statements including IDs. The data frame is then further processed for obtaining the final results and feedback for the user in the output layer as described in Section 3.5. This calculation takes place after the course of the meeting, since it is to intense to be calculated in real-time for most of the tools. Since *TextBlob* and the German version *TextBlob-DE* are both very fast sentiment analysis tools due to the fact of being lexicon based, we calculate additional real-time feedback using the sentiment polarities from both of these tools for the English and German modes respectively. Therefore, we can print the detected sentiment polarities of each collected statement alongside with the statement itself within the *SEnti-Analyzer* as real time feedback for the user (but with only limited reliability, as we only consider the outcome of one tool). We decided to do so by adding a colored square before each statement, with the color green presenting a positive sentiment polarity, the color red presenting a negative sentiment polarity and the absence of color (i.e. white/grey/black, depending on the background properties) presenting a neutral sentiment polarity, for example:

- ☐ This is a neutral statement.
- 🟩 I am very happy.
- 🟥 I am unhappy.

## 3.5 Data Output

On the data frame of statements and corresponding sentiment polarities from all of the tools obtained from the previous classifications layer, we wanted to combine the individually predicted sentiment analysis labels to a combined result for the user. One possibility to achieve this goal is to use majority voting. Majority votes are commonly used in the machine learning domain (e.g., [38, 44, 36, 35, 33, 47]) for all kinds of classification tasks, but the classification categories are usually unordered, while the sentiment polarity classes can be ordered using the ordinal scale: *negative < neutral < positive*. In addition, a majority vote of a tie usually results in of the two tied classes being randomly returned. Assuming our four tools result in two votes for *negative* and two votes for *positive* we get the following outcome:

$$P(negative \mid Majority(negative, negative, positive, positive)) = 0.5$$
$$P(neutral \mid Majority(negative, negative, positive, positive)) = 0.0$$
$$P(positive \mid Majority(negative, negative, positive, positive)) = 0.5$$

where *P* denotes the probability. The majority would would always return *negative* or *positive*, each with a probability of 50%, but never *neutral*. Using the median on the ordinal scale *negative < neutral < positive* on the other hand the result would be *neutral*:

$$Median(negative, negative, positive, positive) = neutral$$

This is correct using the order of *negative < neutral < positive*, as there are as many elements below *neutral* as above (two each here). Therefore, we wanted to use the median to calculate our combined sentiment polarity label out of our four individual labels. However, using the median leads to another problem, assuming we have got two *neutral* predictions and either two *negative* or two *positive* ones:

$$Median(negative, negative, neutral, neutral/positive) = negative + \frac{neutral - negative}{2}$$

$$Median(neutral/negative, neutral, positive, positive) = neutral + \frac{positive - neutral}{2}$$

In the first case we get the arithmetic mean of *negative* and *neutral* as a result and in the second case the arithmetic mean of *neutral* and *positive*. Both are undefined as our ordinal scale *negative* < *neutral* < *positive* misses elements in between *neutral* and *negative* as well as *neutral* and *positive*. We decided to manually override the results for the both cases as *neutral*, as we assume that a developer drawn between the *neutral* and one of the other sentiment polarity classes would lean towards the *neutral* class when annotating sentiment polarity classes to statements. This results in the following definition of *median*$^*$:

$$Median^*(a, b, c, d) = \begin{cases} neutral & \text{if} \quad a = b = negative, c = d = neutral \\ neutral & \text{if} \quad a = b = neutral, c = d = positive \\ neutral & \text{if} \quad a = b = negative, c = neutral, d = positive \\ neutral & \text{if} \quad a = negative, b = neutral, c = d = positive \\ Median(a, b, c, d) & \text{else} \end{cases}$$

where $a, b, c, d \in \{negative, neutral, positive\}$ ordered according to the <-relation. This design choice leads to a higher precision for the negative and positive sentiment polarity classes by nature, as we do not choose these labels in cases where the individual tool predictions are split (i.e. the actual statement might not be positive or negative). Through this we hope to not frequently disturb the project manager or user with false alarms from the tool. The calculated result is also added as the last column of the data frame after the individual labels and outputted as as CSV file for the user. The absolute and relative counts of each resulting sentiment polarity class are computed and outputted for the user as well for a quick first glance evaluation of the course of the meeting.

# 4 Evaluation

In order to evaluate the *SEnti-Analyzer* with respect to its applicability to meetings and its accuracy, we conducted a study. In the following, we present our research objective, the research questions, and the study itself. Our approach basically consists of two steps: (1) the transcription of a meeting and (2) the application of multiple sentiment analysis tool(s).

## 4.1 Research Objective and Research Questions

The main objective of our research is to *analyze the sentiments transported in statements made in a meeting of a software project with sufficient accuracy*. To reach this goal, we developed and evaluated the concept and the corresponding software tool (cf. Section 3) which uses an audio stream of verbal (meeting) communication as input and predicts the polarity of each statement. We formulated the following research questions:

RQ1: How can automatic speech recognition and sentiment analysis be combined to analyze the statements in meetings of a software project?

RQ2: How accurate are the results of our approach by means of matches with predefined labels of scientific authors?

RQ3: How do the automatically produced results differ from the subjective analysis of a computer scientist who is a potential user of the tool?

## 4.2 Procedures to answer RQ1

In the following subsections, we describe the study conducted to answer RQ1 about the applicability of the tool to a software project meeting.

### 4.2.1 The Case Meeting

To obtain proof of concept for the verbal communication mode we tested the *SEnti-Analyzer* on audio extracted from a real student software project meeting. Students in their last year of the bachelor in computer science at Leibniz University Hannover must attend student software project hosted by the Software Engineering Group each fall semester. The professional software development process is brought closer to the students to prepare them for the industry. Groups of five to ten students work together on a software project application, the majority of which are developed for real life local customers (such as the *Hannover Police Department* and the *Hannover Medical School*). The development process of these projects lasts one semester (approx. 15 weeks) including weekly meetings both team-internal and with the customer(s). The project team that participated in the case study worked on the *VirtuHoS*-Project (Virtual House of Software), an application to virtually empathize the feeling of working together in an office with a decentralized development team. The team was tasked to create an editor for drawing a virtual office and creating the underlying semantic structure for further use by other groups using the Java programming language. Due to the ongoing Sars-CoV2 pandemic meetings could only be held virtually. For our case study, we recorded a 33-minute online meeting on 13th January 2021 in which all six team members participated. We collected written consent of each team member allowing us to use the recorded audio files for research purposes and for scientific publications. The team participated voluntarily in the study and the participation had no influence on passing the course, on grades, etc.

### 4.2.2 Data Collection and Pre-Processing

The meeting session was recorded digitally[4]. A single author transcribed and classified the recordings by hand to get reasonable training data for the sentiment classifier proposed by Klünder et al. [26] used in the initial concept of the *SEnti-Analyzer*. The team members used *Discord* as their VoIP service and held the meeting together in a group call. A recording bot was used to record the meeting which enabled a multi-track recording separating each team member from another. To improve the transcription quality we digitally reduced the noise, and adjusted audio levels of all team members equally. Furthermore we limited our test file for obtaining the final results to an representative excerpt of 10 minutes out of the recorded meeting. At the end of the meeting, the team members were asked how they felt about the mood of the meeting (concerning the communication behavior). All team members agreed on the meeting communication being neutral to positive. A second prerecorded meeting from an older iteration of the student software project was also transcribed by hand increase the training set. The complete data set was then fed into the training function of the sentiment classifier proposed by Klünder et al. [26] to find new solutions by hyperparameter search for the included metrics extracted by natural language processing. For this training process, we used 1000 generations in total using an (1 + 1) evolutionary algorithm, thus only introducing one new population per generation and minimizing run-time.

### 4.2.3 Data Analysis

Both transcripts were split into single statements, which were then manually fitted with training labels to create the training data. A special training script loaded the whole data set into the training function of the sentiment classifier proposed by Klünder et al. [26] intending to learn a generalizing model. In total, our training data set consists of 712 manually transcribed and labeled statements, which follow the distribution of 552 (77.5%) neutral, 83 (11.7%) negative and 77 (10.8%) positive statements For the improved version of the *SEnti-Analyzer* no training data was needed as every sentiment analysis classifier uses its own pre-training or lexical data which we assume is superior to our rather small self-collected training data set. However, this gives the initial version [18] the benefit of domain specificity, since it was trained using actual software project meeting data, and not general purpose one. To validate our results, we used Fleiss' $\kappa$ [12] as a statistical measure by comparing the classifications taken by the sentiment classifier proposed by Klünder et al. [26] and the *SEnti-Analyzer* with manual classifications of a human observer for 50 statements transcribed from the case meeting. The 50 statements were manually selected and we choose the ones best transcribed by the speech engine, to be able to classify them manually in a sensible way.

## 4.3 Procedures to answer RQ2

To validate the English mode of operation on the *SEnti-Analyzer* we decided to use two popular pre-labeled data sets, the guideline based *GitHub* gold standard data set by Novielli et al. [41] and the ad-hoc labeled *Stack Overflow* data set by Lin et al. [34]. This is because both data sets have been used frequently in the field of sentiment analysis in software engineering domains (i.e. [57, 56, 42]). As a first step to answer RQ2 on the comparison between the *SEnti-Analyzer* and pre-labeled data sets, we loaded the 7122 statements of the *GitHub* gold standard data set by Novielli et al. [41] and the 1500 statements of the *Stack Overflow* data set by Lin et al. [34] into the *SEnti-Analyzer* using the CSV file mode of operation as described in Section 3.2. We then compared the resulting combined labels of the four sentiment analysis tools calculated as described in Section 3.5 with the author labels of the two data sets individually as well as combined. We did so by computing the overall number of absolute and relative matches (accuracy). In addition, we calculated the shares of mild and severe disagreement cases as proposed by Novielli et al. [41], where mild disagreement means the two differing labels being positive and neutral or negative and neutral, and severe disagreement pointing to a difference between positive and negative prediction and true labels. To validate our results further, we also calculated the Cohen's $\kappa$ [9] statistical measure of observer agreement between the predicted labels of the *SEnti-Analyzer* and the author labels.

---
[4]Due to privacy concerns, we are neither allowed to publish the video recording nor the transcript.

## 4.4 Procedures to answer RQ3

In addition, we have evaluated the data sets used in RQ2 within a survey where we (inter alia) obtained the median labels of (prospective) developers and computer scientist voters for a subset of 48 statements of both data sets in addition to the author labels adding another measure of validation[5].

We received 180 total responses in our survey, including the participants labels for 96 statements extracted from the both data sets mentioned above alongside to demographic characteristics. Out of the 180 total responses we removed 17 data points that answered either question on programming experience or being a computer scientist (*"Would you identify yourself as a computer scientist (e.g., computer scientist student, developer, etc.)?"* or *"Do you have any experience with programming (e.g programming a software, website, an app, etc.)?"*) with a "No" as our target group were persons who are potential team members in a software project team. As a second criterion, we only included data points where participants had at least annotated one of the 96 statements with their perceived sentiment polarity, removing additional 69 data points, making 94 selected data points in total. We calculated the set of median statements to get an overview of the medium perception of a group of developers using the same ordinal scale *negative < neutral < positive* as described in Section 3.5. This way, we calculated a single combined label for each of the 48 statements from each data set, making 96 median human labels in total.

To evaluate the performance of the *SEnti-Analyzer* further, we used our subsets of 48 statements (16-16-16 negative-neutral-positive) each from both data sets [41, 34] where we had the median labels annotated by actual developers from our survey, representing an average developers perception, in addition to the labels annotated by the scientific authors of the data sets [41, 34]. We did so by considering three configurations of predicted and true labels, as illustrated in Figure 2 the first one being the author labels as the true labels and the *SEnti-Analyzer* classifications as the predictions.

The second configuration assumed the human labels as the true labels and the *SEnti-Analyzer* classifications as the predictions. The last configuration compared the author labels as the true labels with the human labels as the predictions. We included this last comparison as a reference point of performance, i.e. to see if we can outperform the average developer. For each of the three configuration (see Fig. 2), we calculated the absolute and relative agreement, as well as Cohen's $\kappa$ [9]. In addition, we computed the precision, recall and $F_1$ metrics for each sentiment class and micro and macro averages for each of our three configurations.

---

[5]Note that we will publish the raw data of the survey as soon as this paper gets accepted. Detailed information on the survey are presented in another paper that is currently under review. At this point, we will present the information required to understand the analysis used in this paper, but not the survey itself.

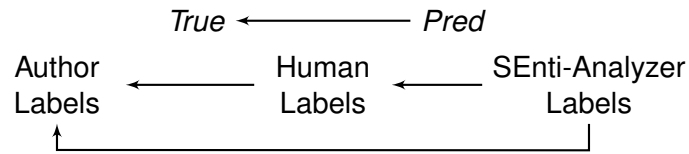

Figure 2: Overview of performed in-depth evaluations to validate the performance of the SEnti-Analyzer. Each arrow present an evaluation with the arrow tip pointing at the set of labels assumed as the true labels

# 5 Results

In the following, we present the results of the data analysis conducted as described before.

## 5.1 Results of the German Sentiment Analysis applied to a meeting

The results for the German mode of operation on the audio of a software project meeting are represented in Table 1. Out of the subset of 50 transcribed statements, 5 (10%) were assigned the label positive by the human observer and none were assigned the label negative, the majority of 45 (90%) statements was labeled as neutral. The sentiment classifier proposed by Klünder et al. [26] identified twice as much statements (10 statements or 20%) with the positive sentiment polarity class, 1 statement (2%) with the negative sentiment polarity class, and 39 statements with the neutral sentiment polarity class, reaching an accuracy of 88% compared to the manual classification and a Fleiss' $\kappa$-score of 0.56, which according to Landis and Koch [32] is considered moderate agreement. The improved version of the *SEnti-Analyzer* using an ensemble of 4 sentiment analysis classification tools as described in Section 3.4 classified 6 statements (12%) with the positive polarity class, 1 (2%) statement with the negative sentiment polarity class and 43 (86%) with the neutral sentiment polarity class. The distribution of sentiment polarity classes of the improved *SEnti-Analyzer* resemble the distribution of sentiment polarity classes of the human observer more than the predictions of the sentiment classifier proposed by Klünder et al. [26]. This is reflected in the slightly increased accuracy of 90%, gaining 2% by the improvements made. The Fleiss' $\kappa$ measure is however slightly lower with 0.5 instead of 0.56, which is likely due to the even more matching classifications in the class neutral and thus a higher chance of a random match considered by Fleiss' $\kappa$.

Table 1: Comparison between classifications taken by the SEnti-Analyzer, the sentiment classifier proposed by Klünder et al. [26], and a human observer

| Classification | Positive | Neutral | Negative | Accuracy | Fleiss' $\kappa$ |
|---|---|---|---|---|---|
| Manual | 5 (10%) | 45 (90%) | 0 (0%) | - | - |
| Klünder et al. [26] | 10 (20%) | 39 (78%) | 1 (2%) | 0.88 | 0.56 |
| SEnti-Analyzer | 6 (12%) | 43 (86%) | 1 (2%) | 0.90 (+0.02) | 0.50(-0.06) |

**Finding 1:** The improvements of the *SEnti-Analyzer* are stand out in a classification which follows the manual classification of a human observer closer than before, and a slight increase in overall accuracy, but a decreased Fleiss' $\kappa$-value.

## 5.2 Results of the Validation on Popular Data Sets

The results of the validation of the sentiment analysis performance capabilities of the newly added English mode of operation of the *SEnti-Analyzer* using the *Stack Overflow* data set by Lin et al. [34] and the *GitHub* gold standard by Novielli et al. [41] and both data sets combined are shown in Table 2.

For the 1500 total statements from the *Stack Overflow* data set [34] the *SEnti-Analyzer* managed to predict 1197 (79.8%) statements correctly. The vast majority of disagreement cases were cases of mild disagreement (i.e. neutral and negative or neutral and positive) with 292 (19.5%) statements, and only a mere 11 (0.7%) cases of severe disagreement (i.e negative and positive). The Cohen's $\kappa$-value reached 0.225 (fair [32]) on the Stack Overflow data set [34]. For the 7122 statement large *GitHub* data set [41] the *SEnti-Analyzer* predicted 5127 (72%) statements correctly. Disagreement was divided in 1872 (26.3%) cases of mild disagreement and 123 (1.7%) cases of severe disagreement. However, the Cohen's $\kappa$-value was much better with 0.577 (moderate [32]).

Table 2: Results of the validation of the SEnti-Analyzer on the Stack Overflow [34] and GitHub [41] data sets

| Data set | Stack Overflow [34] | GitHub [41] | Combined |
| --- | --- | --- | --- |
| Size | 1500 | 7122 | 8622 |
| Balance (Neg-Neu-Pos) | 11.9% - 79.4% - 8.7% | 29.3% - 42.4% - 28.2% | 26.3% - 48.9% - 24.9% |
| Matches (Accuracy) | 1197 (79.8%) | 5127 (72.0%) | 6324 (73.3%) |
| Mild Disagreement | 292 (19.5%) | 1872 (26.3%) | 2164 (25.1%) |
| Severe Disagreement | 11 (0.7%) | 123 (1.7%) | 134 (1.6%) |
| Cohen's $\kappa$ | 0.225 | 0.557 | 0.487 |

**Finding 2:** The overall accuracy of the *SEnti-Analyzer* is higher with the *Stack Overflow* data set [34] (79.8%) than the *GitHub* data set [41] (72%).
**Finding 3:** Cohen's $\kappa$ is higher for the *GitHub* data set [41] (0.557, moderate [32]) than the *Stack Overflow* data set [34] (0.225, fair [32]).

Combining the results of both data sets makes for a total of 8622 statements a number of 6324 (73.3%) correctly classified ones by the *SEnti-Analyzer*. In 2164 (25.1%) cases, the disagreement is only mild, while only being severe in 134 (1.6%) cases. The combined Cohen's $\kappa$-value reaches 0.487 which is considered moderate agreement according to Landis and Koch [32].

**Finding 4:** The *SEnti-Analyzer* manages to be correct in 73.3% of the time. When the *SEnti-Analyzer* does not agree with the author labels, the disagreement is mostly mild (25.1% of the time) and rarely severe (1.6% of the time).

## 5.3 In-depth Validation of Predictions compared to both Author and Human Labels

The measures of matching classifications, agreement, and Cohen's $\kappa$ between the author labels, human labels, and *SEnti-Analyzer* predictions are illustrated in Table 3.

In the original comparison of our case survey as described in Section 4.4 the median human labels from survey participants with software developing backgrounds managed to guess 60 out of the 96 total statement from the data sets [34, 41] correctly. Therefore, the overall agreement measured 62.5% with a Cohen's $\kappa$-value of 0.438. The *SEnti-Analyzer* managed to predict the author labels correctly for 62 out of the 96 statements. The agreement between the *SEnti-Analyzer* predictions and the author labels was 64.6% with a Cohen's $\kappa$-value of 0.469. In addition, when comparing the *SEnti-Analyzer* predictions with the human predictions 61 of the 96 statements were matching. The agreement between human labels and *SEnti-Analyzer* predictions was 63.5% with a Cohen's $\kappa$-value of 0.348.

Table 3: Comparison between the labels of the scientific authors of the data sets (Author), the median labels of the survey participants (Human), and the SEnti-Analyzer (SA)

|               | Matches | Agreement | Cohen's $\kappa$ |
|---------------|---------|-----------|------------------|
| *Author - Human* | 60/96   | 0.625     | 0.438            |
| *Author - SA*    | 62/96   | 0.646     | 0.469            |
| *Human - SA*     | 61/96   | 0.635     | 0.348            |

**Finding 5:** The *SEnti-Analyzer* manages to coincide with both the labels assigned by the scientific authors and the median human labels more than both with each other. That means the *SEnti-Analyzer* predicted the pre-assigned labels of the data sets better than the set of median human labels as well as the perception of the humans better than the pre-assigned labels.

Figure 3 shows a Venn diagram of the overlapping matching classifications between the authors, humans, and the *SEnti-Analyzer* as depicted in Table 3. Between all raters (*SEnti-Analyzer* predictions, authors and median human labels) 46 core statements are always labeled identically. The authors and humans agree on 14 statements additionally, while the authors and the *SEnti-Analyzer* agree on 16 additional statements. The humans and the *SEnti-Analyzer* agree on 15 statements apart from the 46 core statements. Interestingly for 91 out of the 96 overall statements there are always at least two out of three who agree on the sentiment polarity label.

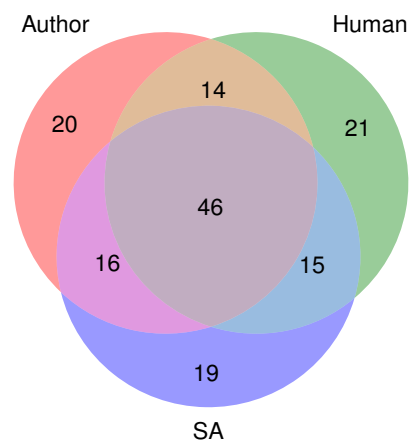

Figure 3: Venn diagram comparing the matching classifications between authors, humans, and the SEnti-Analyzer (SA)

**Finding 6:** The *SEnti-Analyzer*, the authors and the humans all agree on a core set of 46 (47.9%) out of the 96 original statements. An additional 16 statements are classified identically by author and the *SEnti-Analyzer*, while another 15 statements are classified identically by the humans and the *SEnti-Analyzer*.

Table 4 shows the detailed precision, recall and $F_1$-score for each sentiment polarity class as well as the micro and macro average values, to the comparisons of Table 3. The *SEnti-Analyzer* manages to achieve about equal $F_1$-scores ranging from 0.609 to 0.66 among all sentiment polarity classes when compared to the author labels. Precision is usually higher than recall except for the *neutral* sentiment polarity class, which however reaches an almost perfect recall of 0.969. The *SEnti-Analyzer* even achieves a perfect precision of 1 for the negative sentiment polarity class compared to the author labels. Overall the performance seems balanced, and the high precision values for the *positive* and *negative* classes are preferred for sentiment analysis. For the comparison of the *SEnti-Analyzer* with the human labels, the $F_1$-scores varies more, with the *negative* sentiment-polarity class performing worst at 0.429. For the *positive* and *neutral* sentiment polarity classes the *SEnti-Analyzer* reaches an $F_1$-scores of 0.545 and 0.735 respectively. The precision of 0.45 for the *positive* class is lower compared to *neutral* and *negative* (0.694 and 0.643). The recall is the lowest for the *negative* sentiment polarity class (0.321) but significantly higher for the *positive* and *neutral* classes (0.692 and 0.782).

> **Finding 7:** The *SEnti-Analyzer* reaches a high precision for the positive and negative sentiment polarity classes (0.85 and 1 respectively), while maintaining an overall balanced performance when compared to the author labels.

Table 4: Precision (P), recall (R), and $F_1$-score between the labels of the scientific authors of the data sets (Author), the median labels of the survey participants (Human), and the SEnti-Analyzer (SA)

| True - Pred | Polarity Class | P | R | F1 |
|---|---|---|---|---|
| | Positive | 0.770 | 0.313 | 0.444 |
| | Neutral | 0.527 | 0.906 | 0.666 |
| *Author - Human* | Negative | 0.750 | 0.656 | 0.700 |
| | Micro Avg. | - | - | 0.625 |
| | Macro Avg. | 0.682 | 0.625 | 0.604 |
| | Positive | 0.850 | 0.531 | 0.654 |
| | Neutral | 0.500 | 0.969 | 0.660 |
| *Author - SA* | Negative | 1.00 | 0.438 | 0.609 |
| | Micro Avg. | - | - | 0.646 |
| | Macro Avg. | 0.783 | 0.646 | 0.641 |
| | Positive | 0.450 | 0.692 | 0.545 |
| | Neutral | 0.694 | 0.782 | 0.735 |
| *Human - SA* | Negative | 0.643 | 0.321 | 0.429 |
| | Micro Avg. | - | - | 0.635 |
| | Macro Avg. | 0.646 | 0.635 | 0.612 |

## 5.4 Summary

The *SEnti-Analyzer* is capable of performing sentiment analysis on transcribed statements from an audio excerpt of a prerecorded software project meeting. The classifications taken by the *SEnti-Analyzer* mimic the classifications taken by a human and reaches moderate agreement (Fleiss' $\kappa$ = 0.5) [32] for the analyzed German meeting audio file. Furthermore we validated the English performance and text operation mode of the *SEnti-Analyzer* comparing the predictions to the author labels of 8622 statements from two different software engineering domains. The *SEnti-Analyzer* managed to reach an accuracy of 73.3% with only 1.6% of severe disagree-ment, and an overall moderate observer agreement (Cohen's $\kappa$ = 0.487). Further investigations with the median labels of human voters revealed that the *SEnti-Analyzer* managed to classify more statements identically with the authors and humans, than the authors and humans with one another, for a subset of 96 statements with an equal spread of sentiment polarities (according to the authors).

# 6 Interpretation

We presented a concept which is capable of performing sentiment analysis on (prerecorded) meeting audio in real-time, with an in depth analysis using a quartet of sentiment analysis tools after the course of the meeting. Additionally the concept can process textual communication in a common manner. Both modes of operation (textual and verbal communication) can be exercised in English and German language. We tested our tool, the *SEnti-Analyzer*, and showed proof of concept on a real meeting from a student software project. Furthermore, we validated the quality of the classification resulting from the *SEnti-Analyzer* for the English mode of operation using two large pre-labeled data sets from two different software engineering domains. This section discusses the findings with respect to the research question and threats to validity. In the end of this section, we point to future work.

## 6.1 Answering the Research Questions

The findings and results we obtained can be used to answer the research questions as follows:

RQ1: **How can automatic speech recognition and sentiment analysis be combined to analyze the statements in meetings of a software project?** We acquired proof of concept for our approach in the exemplary case study, and therefore successfully combined an automatic speech recognition system with a quartet of existing sentiment analysis tools for English and German. The developed tool, although improved from the initial paper [18] is still an exemplary concept and many further improvements, particularly in speech recognition are indispensable. Nevertheless, we are confident that this approach delivers valuable results and will lead to more productive working environments within software project teams in the future.

RQ2: **How accurate are the results of our approach by means of matches with predefined labels of scientific authors?** The predicted labels of the *SEnti-Analyzer* mostly coincide with the labels predefined by the scientific authors of data sets from different software engineering domains. When disagreement occurs it is usually mild. For a total of 8622 statement the *SEnti-Analyzer* managed to predict the labels correctly in 73.3% of the cases, while disagreement is mild and severe in 25.1% and 1.6% of the time respectively. For a subset of 96 statements with even amounts of neutral, negative and positive statements the *SEnti-Analyzer* achieved 64.6% accuracy while the median labels from survey participants reached 62.5%.

RQ3: **How do the automatically produced results differ from the subjective analysis of a computer scientist who is a potential user of the tool?** For the analysis of statements transcribed from audio by the *SEnti-Analyzer* we reached a moderate agreement ($\kappa$ 0.5) between the automatically produced results and the manual classification of a human observer. However, this outcome may be the result of highly unbalanced distribution of the classified sentiment polarities leading to a high $P_e$-value and thus a lower $\kappa$-value. As for the comparison between the predictions of statements from two software engineering domain data sets the predictions of the *SEnti-Analyzer* coincide with the median perception of survey participant(s) in 63.5% of the time.

## 6.2 Discussion

Summarizing the results of the research presented in this report, we can observe three remarkable findings:

(1) The application of sentiment analysis to verbal communication is as meaningful as the application to textual communication.

(2) A joint sentiment analysis based on established tools leads to promising results, but...

(3) There are huge discrepancies between the predictions of the joint sentiment analysis, pre-labeled data sets, and the median perception of computer scientists.

Despite the fact that we do not have evidence supporting finding (1), this directly emerges from the fact that sentiment analysis itself has potential to support the collaboration in a team [43]. Therefore, it would be meaningful to start exploring the potential and to improve the polarity detection in meetings of software projects in a larger study.

Nevertheless, besides the threats to validity which we discuss in the following section, there are three aspects that should be considered in the context of our study:

### 6.2.1 Transcription Quality is a Bottleneck

Due to the fact that statements are transcribed before the application of sentiment analysis an information loss is possible due to transcriptions errors. While minor transcription are most likely negligible due to factors like lemmatization and stemming applied in sentiment analysis tools, major errors will lead to an overall mis-classified statement. This has to be addressed because for languages other than English, available transcription quality is not yet good enough. Our case study meeting was recorded in fairly high quality and we had the option to remove background noise and adjust audio levels afterwards. We still choose to only classify and compare the 50 best transcribed statements out of 140 manually as others were often not sensible. After all, the German language model used for transcription still had a word error rate of 15.1%. While English models are often quoted with much lower error rates, they are usually much worse for continuous speech and background noise in real world environments. Therefore, within a live meeting setting imperfect audio conditions can interfere with the capability of the speech engine and thus lead to unusable transcripts. We are aware of this and do not recommend industry use of such a concept at the moment. The future will almost certainly bring much better speech recognition technology and lessen the current state bottleneck of the linked system.

### 6.2.2 Difficulty in Labeling Statements and Statement Context

Obaidi and Klünder [43] found multiple authors quoting difficulties in annotating labels while obtaining data sets (e.g., [41, 34, 5, 11, 52, 55, 39, 24, 57, 20, 45, 40]). Everybody has his/her own perception of the sentiment of a statement and, thus, two different people may choose a different sentiment class for the same statement as our comparison between the author and participant labels showed. Therefore, the labels predicted by the *SEnti-Analyzer* that do not coincide with predefined author labels are not necessarily "wrong", because the predefined labels are perhaps also not the ultimate truth. There will always be some edge cases where a clear sentiment polarity cannot be agreed on by all parties. Additionally, since we focus on meeting analysis with our approach the context is non negligible most of the time. While pre-labeled data sets always consider each statement on its own, the same does not go for analyzing meeting data. Statements in development team meetings occur in a chronological sequence by nature and often refer to previous statements by other team members, which has to be taken into consideration. The implementation of the concept of a statement context for the *SEnti-Analyzer* is also imaginable (i.e. through taking the previous sentiment polarities into account when calculating the current). It has already been pointed out by several papers that context was one of the reasons for misclassification of tools or author labels [5, 8, 17]

### 6.2.3 What is the Ultimate Truth?

These thoughts highlight the relevance of a clarification of what we want to predict with sentiment analysis tools. So, in this case, we (researchers) need to ask ourselves whether the comparison with existing data sets is appropriate or whether we should use subjective labels as we did when comparing the tool's output to the median perception of computer scientists. According to the very limited data set used in our study, there are not negligible discrepancies, but how can we be sure that our tool classifies statements correctly? What does "correct" mean? It might be reasonable to rethink the definition of correct labels by trying to calibrate the tool according to a team's need (that is to adjust the tool to the team-setting in which it should be used). If we have a pessimistic team, neither the pre-defined labels in a data set (independent of whether they are labeled ad hoc or using a guideline) nor other data sources might be appropriate to reflect the perceptions of the team members. However, this is what sentiment analysis tools are applied for: to predict how the member(s) of a team perceives a message when he or she receives it. Data sets are required to train the tools to achieve this goal, but at this point in research, it might be meaningful to rethink the procedure and to adjust it according to a specific team's needs.

## 6.3 Threats to Validity

Our case study results are limited to the used sample and cannot be generalized for other meetings. In this section, we summarize the most relevant threats to validity probably impacting our results.

We applied the verbal communication mode of the *SEnti-Analyzer* to a single meeting. The used statistical measure, Fleiss' $\kappa$, may be unsuitable regarding the high number of matching classifications in the class *neutral* by both observers (*SEnti-Analyzer* and classification by hand). This aspect alone had a high influence on the calculated $\kappa$-value and the resulting strength of agreement. Because of the Sars-CoV2 pandemic, and the resulting curfew, the case study meeting had to be held online through VoIP software. Therefore, participants used their computers to attend the meeting. This influenced the experiment environment due to background noise, sounds from other rooms, noise from outside, and other static or interference noises. Delays over the VoIP also showed to be problematic while talking together, e.g., by cutting each other short unintentionally. Additionally, for the comparison of German classification of statements extracted from the meeting audio we picked the 50 statements best transcribed by the *SEnti-Analyzer*, the proportion of statements transcribed with errors is still high, especially on languages other than English, and results are obviously limited by that. However, we are confident that speech-to-text systems will outperform the human performance for all languages eventually, and the *SEnti-Analyzer* is an early concept of what will be capable in the future.

The student software project team recorded in the case study consisted of bachelor degree students in computer science. The prior knowledge about professional software development varied between team members. While some had already worked in private software corporations alongside their studies, others had programming knowledge only consisting of basic programming courses in computer science required to participate in the student software project. Therefore, the team members with less experience also did not use *JIRA* or *GitLab* prior to the software project. These variations may not influence the application of the *SEnti-Analyzer*, but should be taken into account when interpreting the results.

The *SEnti-Analyzer* is currently not capable of differentiating voices, resulting in a transcription that consists merely of a concatenation of all recognized statements, instead of offering dialogue-like structuring. For the transcription to work as intended, it is necessary that only one person talks at a time, or otherwise, the quality of the transcript will be compromised. The overall results are therefore limited by the currently free available transcription technology. Future research needs to focus on these issues.

To validate the performance of the German part of our tool, one of the authors labeled a subset of the statements in the transcript. This introduces a bias as he might have guessed the most likely outcome of the tool. In future research, this bias needs to be resolved. However, generally validating the outcome of the sentiment analysis on the meeting requires future studies and more reliable data to be able to draw conclusions. However, in this first step, we strived to evaluate the applicability. Due to this missing data (e.g., given in a German gold standard), we evaluated the English version of the tool in this paper. This does not reduce the threat of the missing validation of the German version, but is a first step to show that the approach of the tool works. We validated the English version by both comparing the outcome to existing gold standards as well as to the results of a survey with potential team members.

Another possible threat of the approach is the tendency towards "neutral" based on the definition of the calculation of the "median" of the four votings. This tendency originates in the assumption that we want to forecast the perception of a "medium developer" that might be rather neutral than positive or negative (on average). However, for future research, it would be interesting to resolve this issue, e.g., by adjusting or calibrating the tool for a specific team's needs. This adjustment might also include the introduction of weightings for the four tools, e.g., based on the used data or the context to software engineering. It might be meaningful to assign a higher weight to the software engineering specific tools to underline the domain of the language, but future studies need to validate this.

## 6.4 Future Work

To reduce the possible impact of the threats to validity and to increase the reliability of our results, we propose the following steps for future work:

(1) Improve the reliability of the results: First and foremost, a (longitudinal) case study (with computer scientist professionals) and a multi-case study are required to strengthen the results and to evaluate the usefulness of the application for the teams.

(2) Taking facial and the tone expressions into account: As verbal communication is not the only communication used in meetings (rolling eyes or getting loud, e.g., also transport a lot of information), it would be interesting to also consider gestures, facial expressions, etc.

(3) Increase the granularity of the results, e.g., by distinguishing between different categories

as proposed by Klünder et al. [27]. This helps pointing, for example, to destructive behavior which endangers project success by demotivating team members.

(4) In addition to existing gold standard data sets for English, there is a need of gold standards for other languages, as several teams still communicate in their native language (instead using English). Consequently, we strive to develop a data set for German and invite other researchers to also create data sets for other languages.

# 7 Conclusion

Meetings represent a valuable way to communicate within development teams and are essential for every software project. To make software project meetings more effective and productive, and thus increase the overall mood and satisfaction of the project team, automated interaction analysis can be used. As the first step to our long-term research goal of an automated fine-grained interaction analysis, we introduce an approach combining prior interaction analysis research with the latest open source speech recognition achievements. The *SEnti-Analyzer* processes meeting audio by cutting the conversation into single statements and transcribing them in real-time, including the sentiment polarity predictions of a fast lexicon based tool. The *SEnti-Analyzer* classifies the transcribed statements using a quartet of established sentiment analysis tools and returns the overall meeting performance by showing the proportions of the sentiment classes *positive*, *negative*, and *neutral*. This way, project managers, software process engineers, and the like can gain additional informative feedback tracing the course of the meeting with little to no effort. Further actions can be taken due to the given resulting feedback, to improve future meeting behavior and communication.

In a case study, we applied our tool to audio from a real student software project meeting. The *SEnti-Analyzer* delivered results that directly corresponded with the feedback the team members gave themselves. Using our results we could also verify moderate agreement of the classifications taken by the *SEnti-Analyzer* in comparison to a human observer.

Furthermore, the capability of predictions was verified using over 8000 statements from two pre-labeled data sets from differentiating software engineering domains, where the *SEnti-Analyzer* reached moderate agreement and a high number of coinciding classifications. The amount of severe disagreements between the *SEnti-Analyzer*'s predictions and the predefined labels was negligible.

Overall, we propose to keep pursuing research on interaction analysis in software development teams using known sentiment analysis methods and machine learning algorithms to further expand the established concepts by integrating other components such as speech or gesture recognition. Automating tools for ease of use is also an important factor to disseminate interaction analysis in software development.

# Literaturverzeichnis